\documentclass{osa-article}

\journal{osac}

\articletype{Research Article}

\usepackage{lineno}
\usepackage{booktabs}
\newcommand{\ket}[1]{|{#1}\rangle}

\newcommand{\bra}[1]{\langle{#1}|}
\newcommand{\bkt}[2]{\langle{#1}|{#2}\rangle}

\newcommand{\av}[1]{\langle#1\rangle}

\begin{document}
\title{Weak-value technique  for detecting weak magnetic field  based on Faraday magneto-optic effect}

\author{Jing-Hui Huang,\authormark{1} Xue-Ying Duan,\authormark{2,3,4} ,Guang-Jun Wang \authormark{2,3,4} and Xiang-Yun Hu\authormark{1,*}}

\address{\authormark{1}School of Institute of Geophysics and Geomatics, China University of Geosciences, Wuhan 430074, China\\
\authormark{2}School of Automation, China University of Geosciences, Wuhan 430074, China\\
\authormark{3}Hubei Key Laboratory of Advanced Control and Intelligent Automation for Complex Systems\\
\authormark{4}Engineering Research Center of Intelligent Technology for Geo-Exploration, Ministry of Education
}

\email{\authormark{1}jinghuihuang@cug.edu.cn\\
\authormark{*}xyhu@cug.edu.cn} 



\begin{abstract}
We study the amplification of weak magnetic field  with Weak-value technique based on Faraday magneto-optic effect. By using a different scheme to perform the Sagnac interferometer with the probe in momentum space, we have demonstrated the new weak measure protocol to detect the small weak magnetic field by amplifying the phase shift of Faraday magneto-optic effect. At the given the maximum incident intensity of the initial spectrum, the detection limit of the intensity of the spectrometer and the accuracy of detecting weak magnetic field, we can theoretically give the appropriate pre-selection, post-selection and others optical structure before experiment. 
Our numerical results show our scheme with Weak-value technique is effective and feasible to detect weak magnetic field with magnetic field intensity lower than $10^{-10}$ T.
\end{abstract}

\section{Introduction}
\label{intro}

Weak magnetic field measurement sensors are widely used in aerospace, biomedicine, resource detection, earthquake disaster early warning and other fields\cite{2010Advances,2020Magnetic,2020Computational}. There are many ways to measure weak magnetic fields(intensity bellow $10^{-9}$ T) and properties. Sensing methods have been based on the use of induction coils, flux gate magnetometers, magnetoresistive and Hall effect magnetometers, magneto-optical magnetometers, and optically pumped magnetometers\cite{2006Superconducting}.The most sensitive magnetic flux detector is the superconducting quantum interference device(SQUID)\cite{2017An}. This device has demonstrated field resolution at the $10^{-17}$ T level. But such magnetic sensors are bulky and require liquid helium cooling. Equipments and maintenance cost are very expensive, which greatly limits the range of applications. Therefore, the weak magnetic field technology with high sensitivity and low cost are the focal and difficult problem in many fileds.

In recent years, spurred by fundamental studies of quantum phenomena and by developments in precision measurements\cite{2013Weak}, Weak-value technique  has shown great superiority in numerous high precision measurements, such as amplification of angle rotations \cite{PhysRevLett.112.200401}, longitudinal velocity shifts\cite{2013Weak}, frequency shifts\cite{PhysRevA.82.063822},the Goos-Hanchen shift\cite{Santana:19}, rotation velocity\cite{Huang2021}, the photonic spin Hall effect\cite{2021Weak}. The "Weak-value" was first proposed by Aharonov et al\cite{AAV}, where information is gained by weakly coupling the probe to the system. By appropriately selecting the initial and final state of the system, the measurement can be much larger than the eigenvalues of observable.

The Faraday magneto-optic effect(FMOE) is a well known optical phenomenon. If an external magnetic field \textbf{B} is applied to a suitable medium such that there is a field component parallel to the direction of linearly polarized light in the medium, it is found that the direction of polarization of the emergent light has been rotated through an angel $\Phi_{F}$, such that
\begin{eqnarray}
\label{old-FMOEangel}
\Phi_{F}=\int\nolimits_{L} V \textbf{B } \cdot \textbf{dl} 
\end{eqnarray}
where $L$ is the path traversed by the light in the medium and $V$ is the Verdet constant. Over the past decade, a considerable number of new optical current sensors that exploit the Faraday magneto-optic effect in optical fiber or bulk glass to measure large currents at high voltage had been developed and reported\cite{1993Faraday,2003Review}. In these works, the direction of the magnetic field is rotational due to the Ampere circuital theorem, therefore, the integral of magnetic field along a closed optical loop around the current could be measured. When taking the weak scalar magnetic field into account, it is hard to detect small angel $\Phi_{F}$ by using Eq.(\ref{old-FMOEangel}) directly. Inspired by the successes of high precision measurements with Weak-value  technique , we further study in this paper Weak-value technique for detecting weak magnetic field  based on the  Faraday magneto-optic effect.
\begin{figure*}[t]
 \vspace{-0.2cm}
\centering
\resizebox{0.99\textwidth}{!}{%
  \includegraphics{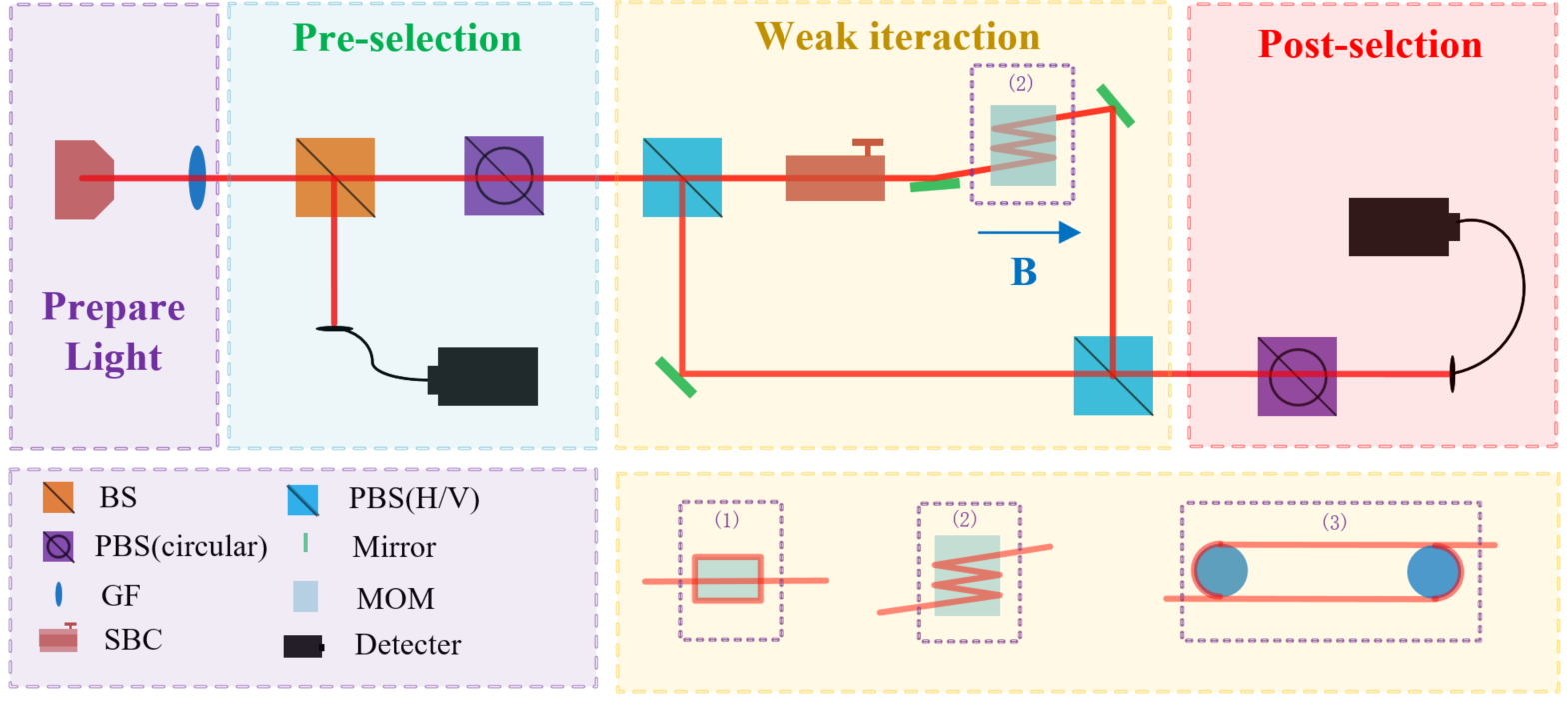}
}
\caption{Schematic of the Weak-value technique for detecting weak magnetic field  based on the  Faraday magneto-optic effect. The light source is shaped by a Gaussian filter(GF). BS is the beam splitter. PBS(circular) is the polarizing beam splitter which changes the light into circular polarization. PBS(H/V) is the polarizing beam splitter which separates the light into horizontal( $\ket{H}$) component and vertical( $\ket{V}$) component. SBC is the Soleil-Babinet compensator which produces certain retardation between $\ket{H}$ and $\ket{V}$. The Magneto-optic material is a special optical component which is made of specific substance with the Verdet constant $V$.}
\label{fig:interferometer}       
\end{figure*}

The rest of this paper is organized in the following way.
In Section 2, we present a scheme of Weak-value technique for detecting weak magnetic field  based on the  Faraday magneto-optic effect, and the numerical results are shown in Section 3. Finally, in Section 4, we give the conclusion about the work. Throughout this paper we adopt the unit $\hbar =1$.

\section{Weak-value technique for detecting weak magnetic field }
\label{sec:deltz}

In this section, we propose a new scheme of Weak-value technique for detecting weak magnetic field  based on the  Faraday magneto-optic effect in the frequency domain, which was reported to be superior to that in the time domain in high precision measurements\cite{PhysRevLett.105.010405}. In our scheme, the direction of the weak magnetic field is unidirectional and our design goal is to detect a magnetic field of magnitude $10^{-9}$ T. The weak measurement is characterized by state preparation, a weak perturbation, and postselection.
We prepare the initial state $\ket{\phi_{i}}$ of the system and $\ket{\psi_{i}}$ of the probe. 
After a certain interaction between the system and the probe, we postselect a system state $\ket{\phi_{f}}$ and obtain information about a physical quantity $\hat{A}$ from the probe wave function by the weak value
\begin{eqnarray}
\label{weak_value}
A_{w}:=\frac{\bra{\phi_{f}}\hat{A}\ket{\phi_{i}}}{\bkt{\phi_{f}}{\phi_{i}}},
\end{eqnarray}
which  can generally be a complex number.
More precisely, the shifts of the momentum in the probe wave function are given by the imaginary parts of the weak value Im$[A_{w}]$. 
We can easily see from Eq. \ref{weak_value} that when $\ket{\phi_{i}}$ and $\ket{\phi_{f}}$ are almost orthogonal, the absolute value of the weak value can be arbitrarily large.
This leads to the weak-value amplification, as we will explain it below.

The schematic diagram of the system is shown in Fig. \ref{fig:interferometer}. The incident light with central wave-length of $\lambda_{0}$ (corresponding to momentum $p_{0}$) passes through a Gaussian filter and the beam splitter. The purposes of the BS are to record the initial spectrum and to compare this spectrum to the final one. According to the previous work\cite{PhysRevLett.105.010405}, the weak measurement scheme based on circular polarizations as pre-selection and post-selection shows a higher precision than that based on linear polarizations.
Thus we select the initial polarization  and the final polarization state of system  using PBS(circular):
\begin{eqnarray}
\label{inter_sy_initial}
\ket{\phi_{i}}= sin(\beta +\frac{\pi}{4})\ket{H}+ icos(\beta +\frac{\pi}{4})\ket{V})
\end{eqnarray}
\begin{eqnarray}
\label{inter_sy_final}
\ket{\phi_{f}}= isin( \frac{\pi}{4})e^{i\varphi}\ket{H}
+ cos(\frac{\pi}{4})e^{-i\varphi}\ket{V}
\end{eqnarray}
where $\ket{H}$, $\ket{V}$ represent the horizontal and vertical polarization, $\beta$ is corresponding the pre-selection. The phase shift ${\varphi}$ is produce by the SBC, Magneto-optic material(MOM) and optical path difference(OPD) between $\ket{H}$ and $\ket{V}$:
\begin{eqnarray}
\label{inter_sy_phase_difference}
\varphi= \varphi_{\rm SBC}+\varphi_{\rm MOM}+\varphi_{\rm OPD}
\end{eqnarray}

According to the Eq.\ref{old-FMOEangel}, it is effective
way to enhance the Faraday rotation $\varphi_{\rm MOM}$ by increase the length of the interaction between light and MOM. The phase shift $\varphi_{\rm MOM}$ is generated by FMOE and we propose three effective schemes in Fig. \ref{fig:interferometer} to detect weak magnetic field: (1)$\varphi_{\rm MOM}=V \textbf{B }D$, where $D$ is the length of the material in the direction of the magnetic field. (2) $\varphi_{\rm MOM}= N V \textbf{B }D$, where $N$ is the number of times light is reflected in the MOM. In this case, the Faraday rotation $\varphi_{\rm MOM}$ is enhanced N times due to the nonreciprocal of FMOE. (3) $\varphi_{\rm MOM}= M V \textbf{B }L$, where $M$ is the number of rings in an optical fiber and $L$ is the circumference of the optical fiber.

The phase shift ${\varphi_{SBC}}$  is produced by the SBC, which can modulate the initial phase shift ${\varphi}$. Before detecting the weak magnetic filed, the "Weak interaction" part of Fig. \ref{fig:interferometer} places in a magnetically shielded environment, then the SBC can be adjusted to satisfy the condition $\varphi= \varphi_{\rm SBC}+\varphi_{\rm OPD}=0$, in that condition there is no shift between the initial and final probe. After the modulation process  complete, the total phase shift ${\varphi}$ is only depended on the $\varphi_{\rm MOM}$.

In our scheme, the observable $\hat{A}$ satisfies:
\begin{eqnarray}
\label{sy_observable}
\hat{A}= \frac{1}{2}(\ket{H} \bra{H}-\ket{V} \bra{V})
\end{eqnarray}

And the weak-value can be calculated by:
\begin{eqnarray}
\label{weak_value1}
A_{w}=\frac{\bra{\phi_{f}}\hat{A}\ket{\phi_{i}}}{\bkt{\phi_{f}}{\phi_{i}}}
=\frac{sin(\varphi)sin(\beta)+icos(\varphi)cos(\beta)}
{sin(\varphi)cos(\beta)+isin(\beta)cos(\varphi)},
\end{eqnarray}

After weak correlation  and postselection, the probe wave function in the momentum space becomes 
\begin{eqnarray}
\bkt{p}{\psi_{f}} &=& \bkt{\phi_{f}}{\phi_{i}}e^{-igA_{w}\hat{p}}\bkt{p}{\psi_{i}}  \\
&=& [{sin(\varphi)cos(\beta)+isin(\beta)cos(\varphi)}] e^{gp \rm Im(A_{w}) }\bkt{p}{\psi_{i}} \nonumber
\end{eqnarray}
and therefore, its absolute value squared is given by:
\begin{eqnarray}
\label{eq_final_probe1}
|\bkt{p}{\psi_{f}}|^{2} =|\bkt{\phi_{f}}{\phi_{i}}|^{2}
 \times e^{2pg\rm Im(A_{w})} |\bkt{p}{\psi_{i}}|^{2}
\end{eqnarray}
where $|\bkt{\phi_{f}}{\phi_{i}}|^{2}$ is the probability to pass the post-selection
\begin{eqnarray}
\label{eq_final_probe2}
|\bkt{\phi_{f}}{\phi_{i}}|^{2} =sin^{2}(\varphi)cos^{2}(\beta) 
+ sin^{2}(\beta)cos^{2}(\varphi)
\end{eqnarray}

In our weak measurement protocol, we can get the shift of the center wavelength $\lambda_{0}$ from  the relationship $\Delta\av{\hat{p}}= 2g W^{2} Im A_{w}$ with $g=2 \pi / p_{0}$ \cite{1990Properties,2018Optical,Li:16}, $p_{0}$ corresponds to the center wavelength $\lambda_{0}=2 \pi / p_{0}$.
\begin{eqnarray}
\label{inter_delt_lambda}
\delta \lambda_{0}=-\frac{4 \pi (\Delta \lambda)^{2}}{\lambda_{0}} \rm Im A_{w}
\end{eqnarray}

It is noteworthy that under actual experimental conditions, it is efficient and convenient to record the spectrum and fitting the central wavelength of the spectrum with the Gauss function. Finally, the relationship of the shift of the central wavelength $\delta \lambda_{0}$ and the weak magnetic field can be obtained from Eq. (\Ref{inter_delt_lambda}).   
In this paper, the results of our simulation experiment will be shown in the next section.

\section{Numerical result and discussion}

\begin{figure}[htp!]
	\centering
	\centerline{\includegraphics[scale=0.99,angle=0]{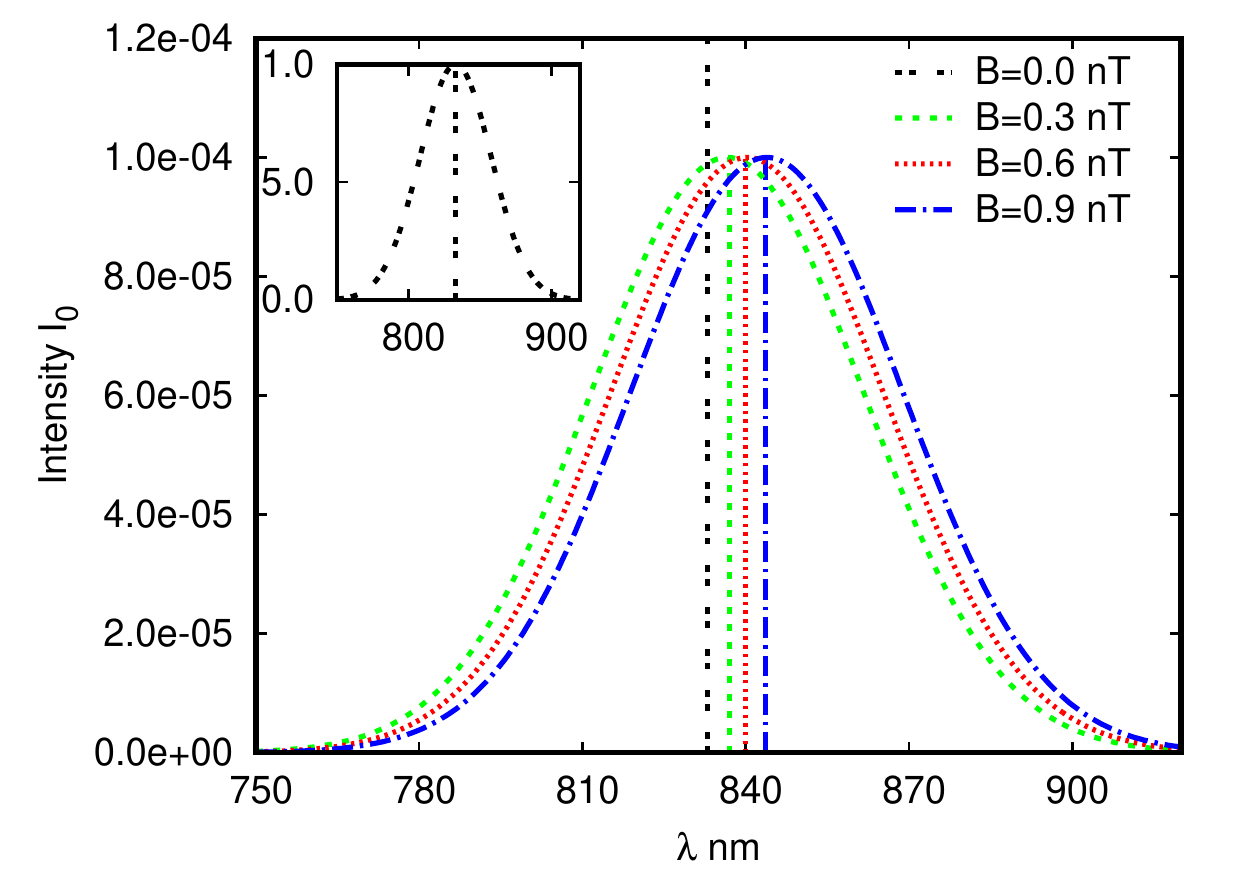}}
\vspace*{0mm} \caption{\label{result_numberical1}The central wavelength shifts of our simulation experiment with pre-selection angle $\beta$ = 0.010 rad in different weak magnetic fields. The vertical lines represent the central wavelengths of the corresponding Gaussian spectrum. }
\end{figure}

\begin{figure}[htp!]
	\centering
	\centerline{\includegraphics[scale=0.99,angle=0]{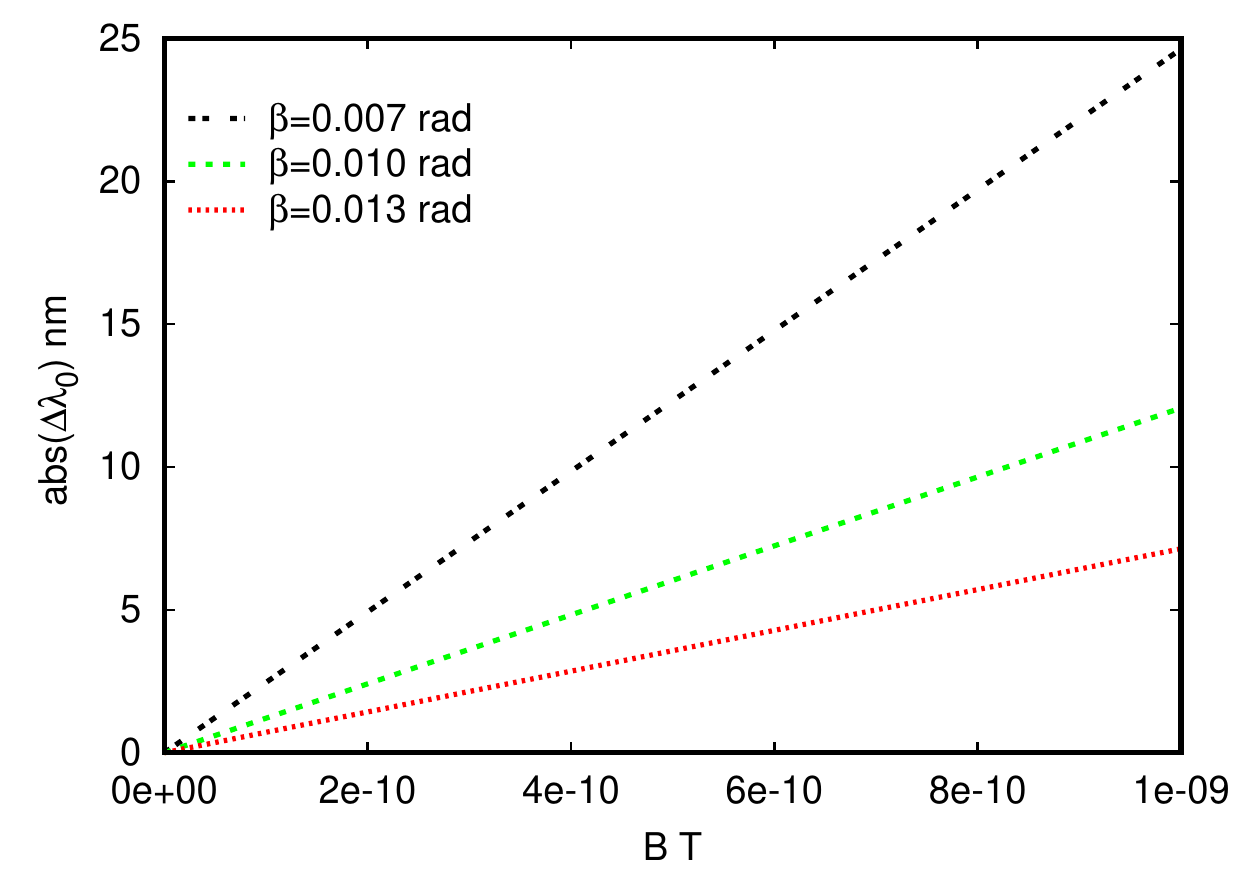}}
\vspace*{0mm} \caption{\label{result_numberical2}The results of our simulation experiment with different pre-selection angles and  in the different magnitude of the magnetic field.}
\end{figure}

In what follows, we fix a specific experimental setup, that is, the given chosen pre-selected state, post-selected state and the initial probe in momentum space, so that we can calculate the weak value and the shift of the center wavelength of the probe before the experiment. In our numerical model, we take the initial probe wave function in Gaussian form:
\begin{eqnarray}
\label{gaussian}
{\Gamma}_{i}(p)=|\bkt{p}{\psi_{i}}|^{2}={\Gamma}_{i}(\lambda) =I_{0}e^{-(\lambda-\lambda_{0})^{2}/W^{2}},
\end{eqnarray}
where $\lambda_{0}=833$ nm is the central wavelength of the initial spectrum and $W^{2}=(\Delta \lambda)^{2}=50$ nm is the variance of the initial spectrum. Considering the implementation difficulty of the scheme, we thought the third option (3) in the Fig.\ref{fig:interferometer} is superior to others, since the optical fiber can be wound over very long lengths. Besides, the effective Verdet constant of the terbium-doped fiber in the work\cite{2010Compact} is measured to be –32.1 ± 0.8 rad/(Tm), which is 27 times larger than that of silica fiber. Therefore, in  our numerical simulation, we take the third option (3) in the Fig.\ref{fig:interferometer} to enhance the Faraday rotation. In particular, we take the total length of the optical fiber $ML$ = 1000 m and the Verdet constant $V$ = 32 rad/(Tm). Then, by using the Eqs. (\ref{eq_final_probe1}) and (\ref{inter_delt_lambda}), we obtain the final probe function and the shifts of the center wavelength with different values of the pre-selection angle $\beta$. The simulation results are shown in Fig. \ref{result_numberical1}, Fig. \ref{result_numberical2} and Table. \ref{tab:1}.

\begin{table}[t]
\centering
\caption{Parameters and the numerical values of our obtained results:  $\beta $ $[rad]$ is corresponding to the pre-slection ; k=$d| \Delta \lambda_{0})|/d B$ [nm/T] is the sensitivity of our scheme and the absolute value of the slope in the Fig. \ref{result_numberical2}.}
\label{tab:1}       
\begin{tabular}{lllll}
\toprule
 $\beta  $ rad&  k nm/T &  $|\bkt{\phi_{f}}{\phi_{i}}|^{2}$ \\
\noalign{\smallskip}\hline\noalign{\smallskip}
0.007 & 2.46 $\times10^{10}$ & 4.9 $\times10^{-5}$\\
0.010 & 1.20 $\times10^{10}$ & 1.0 $\times10^{-4}$\\
0.013 & 0.71 $\times10^{10}$ & 1.6 $\times10^{-4}$\\

\toprule\\
\end{tabular}
\vspace*{-0.6cm}  
\end{table}

Fig. \ref{result_numberical1} shows the shifts of the Gaussian spectrum with pre-selection angle $\beta$ = 0.010 rad in different weak magnetic fields. By fitting each central wavelength of the Gaussian spectrum in different weak magnetic fields, we obtain the shifts of the Gaussian spectrum due to the change in the magnitude of the magnetic field.
More specifically, the shift of the central wavelength increases as the magnetic field strength increases.

Fig. \ref{result_numberical2} and the Table. \ref{tab:1} show the sensitivity of our scheme with different pre-selection angles, which  corresponds to the slope of the curve. Our numerical results show the smaller the $\beta$ is , the larger the amplification(corresponding to the slope) is . On the other hand, the smaller $\beta$ leads to the lower probability $|\bkt{\phi_{f}}{\phi_{i}}|^{2}$ to detect the post-selection. Therefor, the value of $\beta$ cannot be infinitesimally small due to the measurement limit of the detection instrument.

We have made the first step towards the numerical study of Weak-value technique  for detecting weak magnetic field  based on Faraday magneto-optic effect. Our numerical results show our scheme can effectively measure weak magnetic fields with magnetic field intensity lower than $10^{-10}$ T.

\section{Conclusion}
In conclusion, we use Weak-value technique to probe weak magnetic fields with magnetic field intensity lower than $10^{-10}$ T based on Faraday magneto-optic effect. By choosing the appropriate pre-selection, post-selection, the initial probe in the momentum space and the Magneto-optic material, we obtain the relationship between the shifts of the center wavelength and weak magnetic field. This amplification effect can’t be explained by classical wave interference \cite{Heiner2007Broad,1991Optical}, due to the statistical feature of pre-selection and post-selection with disturbance\cite{PhysRevLett.113.120404}. 

Before performing specific experiments, our numerical results have confirmed that our scheme can effectively measure weak magnetic fields with magnetic field intensity lower than $10^{-10}$ T. Besides, in order to detect the weaker magnetic field, it is effective and feasible to increase the maximum incident intensity $I_{0}$ of the initial spectrum, improve the measurement precision of the spectrometer, increase the length of the fiber ring and look for magneto-optic materials with larger Verdet constant. 

On the other hand, at the given the maximum incident intensity $I_{0}$ of the initial spectrum, the detection limit of the intensity of the spectrometer and the accuracy of detecting weak magnetic fields, we can theoretically give the appropriate post-selection, pre-selection and others optical structure before experiment. In addition, the relevant optical experiments are taking in progress.
%
%
\section{Authors contributions}
All the authors were involved in the preparation of the manuscript.
All the authors have read and approved the final manuscript.

\section*{Acknowledgements}
This study is financially supported by the National Key R$\&$D Program of China (No. 2018YFC1503705).
We acknowledge financial support from National Natural Science Foundation of China (Grants No. G1323519204). The project was supported by the Fundamental Research Funds for National Universities,China University of Geosciences(Wuhan).

\bibliography{sample}

\end{document}